\documentclass[12pt]{article}

\usepackage{natbib}

\usepackage{acro}
\usepackage{amsmath}
\usepackage{amssymb}
\usepackage{amsthm}
\usepackage{bm}
\usepackage{booktabs}
\usepackage{fullpage}
\usepackage{graphicx} 
\usepackage{hyperref}
\usepackage{bbm}
\usepackage{cleveref}
\usepackage[usenames,dvipsnames]{xcolor}
\usepackage{xspace}

\DeclareAcronym{PrO}{short=PrO, long=predictively oriented}
\DeclareAcronym{EKF}{short=EKF, long=extended Kalman filter}
\DeclareAcronym{KLD}{short=KLD, long=Kullback--Leibler divergence}
\DeclareAcronym{DoC}{short=DoC, long=difference-of-convex}
\DeclareAcronym{MSE}{short=MSE, long=mean square error}

\definecolor{EKF}{RGB}{32, 178, 170}
\definecolor{EKF-B}{RGB}{255, 140, 0}
\definecolor{EKF-IW}{RGB}{220, 20, 60}
\definecolor{OGD}{RGB}{138, 43, 226}
\definecolor{WLF-IMQ}{RGB}{30, 144, 255}
\definecolor{WLF-MD}{RGB} {229, 193, 0} 
\definecolor{Hub-EnKF}{RGB}{0, 116, 0}
\definecolor{WS-EnKF}{RGB}{198, 122, 55}
\definecolor{tabblue}{RGB}{31, 119, 180}
\definecolor{taborange}{RGB}{255, 127, 14}
\definecolor{mediumorchid}{HTML}{BA55D3}

\newcommand{\mWlfMd}{{\color{WLF-MD} \texttt{\textbf{WoLF-TMD}}}\xspace} %
\newcommand{\mWlfImq}{{\color{WLF-IMQ} \texttt{\textbf{WoLF-IMQ}}}\xspace}

\newcommand{\mkfExtendedIterated}{{\color{EKF} \texttt{\textbf{I-EKF}}}}

\newcommand{\mkf}{{\color{EKF} \texttt{\textbf{KF}}}\xspace}
\newcommand{\mkfExtended}{{\color{EKF} \texttt{\textbf{EKF}}}\xspace}

\newcommand{\mAgamenoniExtended}{{\color{EKF-IW} \texttt{\textbf{EKF-IW}}}\xspace}
\newcommand{\mWangExtended}{{\color{EKF-B} \texttt{\textbf{EKF-B}}}\xspace}
\newcommand{\mPrO}{{\color{mediumorchid}\texttt{\textbf{EKF-PrO}}}\xspace}
\newcommand{\mPrOKF}{{\color{mediumorchid}\texttt{\textbf{KF-PrO}}}\xspace}

\newcommand{\KFmean}{{\color{EKF} \text{\normalfont \texttt{\textbf{m}}}_k} \xspace}
\newcommand{\KFcov}{{\color{EKF} \text{\normalfont \texttt{\textbf{P}}}_k} \xspace}
\newcommand{\KFgain}{{\color{EKF} \text{\normalfont \texttt{\textbf{K}}}_k} \xspace}

\DeclareMathOperator*{\argmin}{arg\,min}

\newtheorem{remark}{Remark}
\newtheorem{example}{Example}
\newtheorem{definition}{Definition}
\newtheorem{proposition}{Proposition}

\usepackage{algorithm}
\usepackage{algpseudocode}

\algrenewcommand\algorithmiccomment[1]{\hfill$\triangleright$ #1}
\algrenewcommand\algorithmicrequire{\textbf{Input:}}
\algrenewcommand\algorithmicensure{\textbf{Output:}}

\usepackage{tikz}
\usetikzlibrary{arrows.meta}
\usetikzlibrary{positioning}

\newif\ifshowextra

\showextrafalse

\title{Predictively-Oriented Kalman Filtering}
\author{Zheyang Shen$^1$, Gerardo Duran-Martin$^2$, Chris. J. Oates$^{1}$ \\
\small $^1$Newcastle University \\
\small $^2$University of Oxford 
}

\begin{document}

\maketitle

\begin{abstract}
This paper presents a post-Bayesian approach to online filtering in nonlinear state-space models, capable of avoiding over-confident inferences in settings where either the dynamical model, the measurement model, or \emph{both}, could be misspecified.
This is addressed using \emph{predictively oriented} (PrO) posteriors, an emerging paradigm in which learning (i.e., posterior concentration) occurs if and only if the overall model is well-specified, without strict adherence to Bayes' theorem. 
As the characterisation of PrO posteriors is challenging, 
our main technical contribution is a fast approximate linear-Gaussian update procedure, analogous to an (iterated) extended Kalman filter.
The methodology, which we call \mPrO,
has no tunable hyper-parameters and has a computational cost comparable to that of existing filtering methods.
Performance is empirically assessed on a range of linear and non-linear applications, in which the state-space model is systematically misspecified.
\end{abstract}

\section{Introduction}
\label{sec: intro}

Our setting is a state-space model
\begin{align}
    x_k & = f(x_{k-1} , \epsilon_k) \label{eq: SS1} \\
    y_k & = h(x_k , \xi_k) \label{eq: SS2}
\end{align}
where the noise terms $(\epsilon_i)_{i \in \mathbb{N}}$ and $(\xi_i)_{i \in \mathbb{N}}$ are independent. 
The functions $f$ and $h$ may be nonlinear and only the $y_k$ are observed.
The \emph{filtering} task refers to the inference of the state variable $x_k$ given the observations $y_{1:k}:=(y_i)_{i=1}^k$.
Note that this set-up also allows for $f$ and $h$ to depend on one or more unknown (static) parameters $\theta$, which can be cast as additional components of the state variable $x_k$ and held constant under the dynamic model \eqref{eq: SS1}.
State-space models of the form \eqref{eq: SS1} and \ref{eq: SS2} arise in time series analysis \citep{durbin2012time}, tracking in computational vision \citep{chen2011kalman}, and sensor fusion in robotics \citep{thrun2002probabilistic}.

Since the latent state $x_k$ may not be strongly constrained from the observations $y_{1:k}$, a Bayesian approach to quantify uncertainty in $x_k$ is natural in the filtering context.
For linear $f$, $h$, and Gaussian $\epsilon_k$, $\xi_k$ with known mean and covariance, the \emph{Kalman filter} (\mkf)
is an efficient online algorithm for exact calculation of the posterior (or \emph{filtering}) distributions $p(x_k| y_{1:k})$ \citep{kalman1960}.
For nonlinear $f$, $h$, algorithms such as the \emph{extended} Kalman filter (\mkfExtended) exploit local linearisation to approximate the filtering distribution while retaining the efficient online nature of the \mkf.
A textbook treatment of algorithms for Bayesian inference in the state-space model in \eqref{eq: SS1} and \eqref{eq: SS2} can be found in \citet{Sarkka23}.

The focus of this paper is the  filtering problem in settings where
\emph{the true data-generating process is not an instance of the state-space model} in \eqref{eq: SS1} and \eqref{eq: SS2}.
In such settings Bayesian inference, and the various numerical methods that approximate Bayesian inference (e.g. \mkfExtended), can produce inferences and predictions that are \emph{over-confident} because they are not capable of recognising and adapting to model misspecification. 
Accordingly, several solutions have been proposed to mitigate this effect:

\paragraph{Misspecification of the Measurement Error Model}

One common cause of misspecification for the measurement error model \eqref{eq: SS2} is the presence of \emph{outliers} in the dataset \citep{liu2020data}.
Accordingly several potential remedies have been proposed, including hierarchical measurement models; 
see e.g., \citep{ting2007, piche2012, Agamennoni2012, nurminen2015, Huang2016, wang2018},
or the use of robust (e.g. Huber)
loss functions in place of the usual negative log-likelihood \citep[e.g.][]{boncelet1983, karlgaard2015nonlinear, das2023robust}.
The latter can be considered as \emph{generalised} Bayesian methods \citep{bissiri2016general,knoblauch2019generalized}, and were further developed from this perspective in \citet{boustati2020generalised,duran2024outlier,duran2025unifying}.
In particular, by departing from the traditional Bayesian update it is possible to down-weight the influence of outliers as they are encountered.
That is, if an observed datum $y_k$ is far into the tail of the data predictive distribution based on $y_{1:k-1}$, then the effect of $y_k$ on inferences and predictions can be manually discounted. 
This idea enabled \citet{duran2024outlier} to rigorously establish outlier-robustness properties of their generalised Bayes method, called the \emph{weighted observation likelihood filter} (WoLF).
However, while generalised Bayesian methods are capable of ignoring occasional outliers, they do not provide a solution to \emph{systematic} model misspecification; if all data were deemed to be outliers, then all data would be ignored. 

\paragraph{Misspecification of the Dynamical Model}
An alternative form of missppecification arises
from an incomplete knowledge of the latent dynamics \eqref{eq: SS1}.
This occurs, for example, when tracking high-speed maneuvering objects,
when a system undergoes unmodelled phase transitions (see e.g., top plot in Figure \ref{fig: example}),
or when the transition function is incorrectly parametrised.
In such cases, the resulting observations may exhibit abrupt jumps,
or the estimation error can increase over time---a phenomenon commonly referred to as \emph{divergence} \citep{schlee1967divergencekf,fitzgerald2003divergencekf}.
To address this form of misspecification, various methods that admit closed-form (including fixed-point) solutions have been proposed.
This includes approaches that assume heavy-tailed noise in the observation and latent space
\citep{roth2013studentfilter,huang2017ststfilter},
variational Bayesian methods for estimating unknown system dynamics \citep{roth2017studentfilter,huang2020slidekf,vilmarest2024viking}, as well as heuristics that perform \textit{covariance inflation} for adaptation
\cite{mehra2003approaches,kelly2002ridgekf,chang2023low}.
However, these approaches typically rely on explicit parametrisations of the latent noise process $(\epsilon_i)_{i \in \mathbb{N}}$, or specification of hyper-parameters, requiring \emph{a priori} knowledge of how the dynamical model might be misspecified.
Bayesian deep learning offers a point of contrast, enabling dynamics to be \emph{learned} \citep[e.g.][]{dahan2025bayesian}, but at the expense of introducing hyper-parameters and substantially increased computational overhead.

\medskip

In summary, the state-of-the-art for closed-form Bayesian filtering under model misspecification mainly focuses on either defining a more sophisticated hierarchical state-space model
and developing efficient computational methods to facilitate Bayesian inference,
or down-weighting the contribution from particular observations $y_k$ within a generalised Bayesian framework.
Neither approach is ideal; we are typically unable to write down a well-specified state-space model, and typical (generalised) Bayesian posteriors do not provide calibrated uncertainty when the statistical model is misspecified, e.g. as demonstrated in the recent work of \citet{shenprediction}.
In the filtering context, a collapse of uncertainty causes overconfident inferences and predictions when the state-space model is misspecified, as illustrated in the first two columns of \Cref{fig: example}.

\subsection{Our Contributions}

To deal with generic model misspecification in \eqref{eq: SS1} and \eqref{eq: SS2}, we contend that an alternative to (generalised) Bayesian inference is required; rather than attempting to fit a more accurate model in a Bayesian framework in the hope that the resulting posterior predictions will be well-calibrated, we propose instead to target the \emph{predictive} performance of the original state-space model. 
Our inspiration comes from deep learning, where the pitfalls of Bayesian inference are explicitly acknowledged and predictively oriented techniques such as posterior tempering are routinely used \citep{kuleshov2018accurate,wenzel2020good}. 
However, \emph{post hoc} techniques such as posterior tempering introduce additional hyper-parameters, a problem which we seek to avoid.

To achieve this we pursue a promising and emerging research direction in post-Bayesian methodology, which was termed \emph{predictively oriented} (\acs{PrO}) posteriors in \citet{mclatchie2025predictively}.
\Ac{PrO} posteriors seek a distribution over latent variables for which the implied data predictive distribution most closely matches the observed dataset.
This represents a fundamental departure from Bayesian and generalised Bayesian methods.
For example, in the case of a static model of observations parametrised by $\theta$,
the \ac{PrO} posterior will concentrate around a single ``best'' parameter $\theta_\star$ if and only if the model is well-specified.
In contrast, both Bayesian and generalised Bayesian posteriors concentrate around $\theta_\star$ irrespective of whether the model is well-specified \citep{miller2021asymptotic}; this leads to a problematic \emph{collapse of uncertainty} when the model is misspecified.

The main technical challenge in developing a \ac{PrO} posterior for state-space models is that \ac{PrO} posteriors are only implicitly defined, in contrast (generalised) Bayesian posteriors for which one can express the posterior density as the product of the prior density and a (generalised) likelihood.
Although \ac{PrO} posteriors can be approximated to arbitrary precision using techniques such as gradient flows and mean-field Langevin dynamics,
these techniques are computationally intensive and therefore unsuitable in the state-space context \citep{chazal2025computable}.
Our main technical contribution in \Cref{sec: pro state space}
is therefore a fast approximate update for \ac{PrO} posteriors in linear (or linearised) state-space models, analogous to how the \mkfExtended enables fast approximate inference in the Bayesian framework.

Through detailed empirical assessment in \Cref{sec: experiments},
we demonstrate that \mPrO out-performs existing robust filtering methods when the state-space model is systematically misspecified.
A discussion of our findings is contained in \Cref{sec: discuss}.

\section{Methodology}
\label{sec: pro state space}

\begin{figure*}[t!]
\centering
\includegraphics[width=0.9\textwidth]{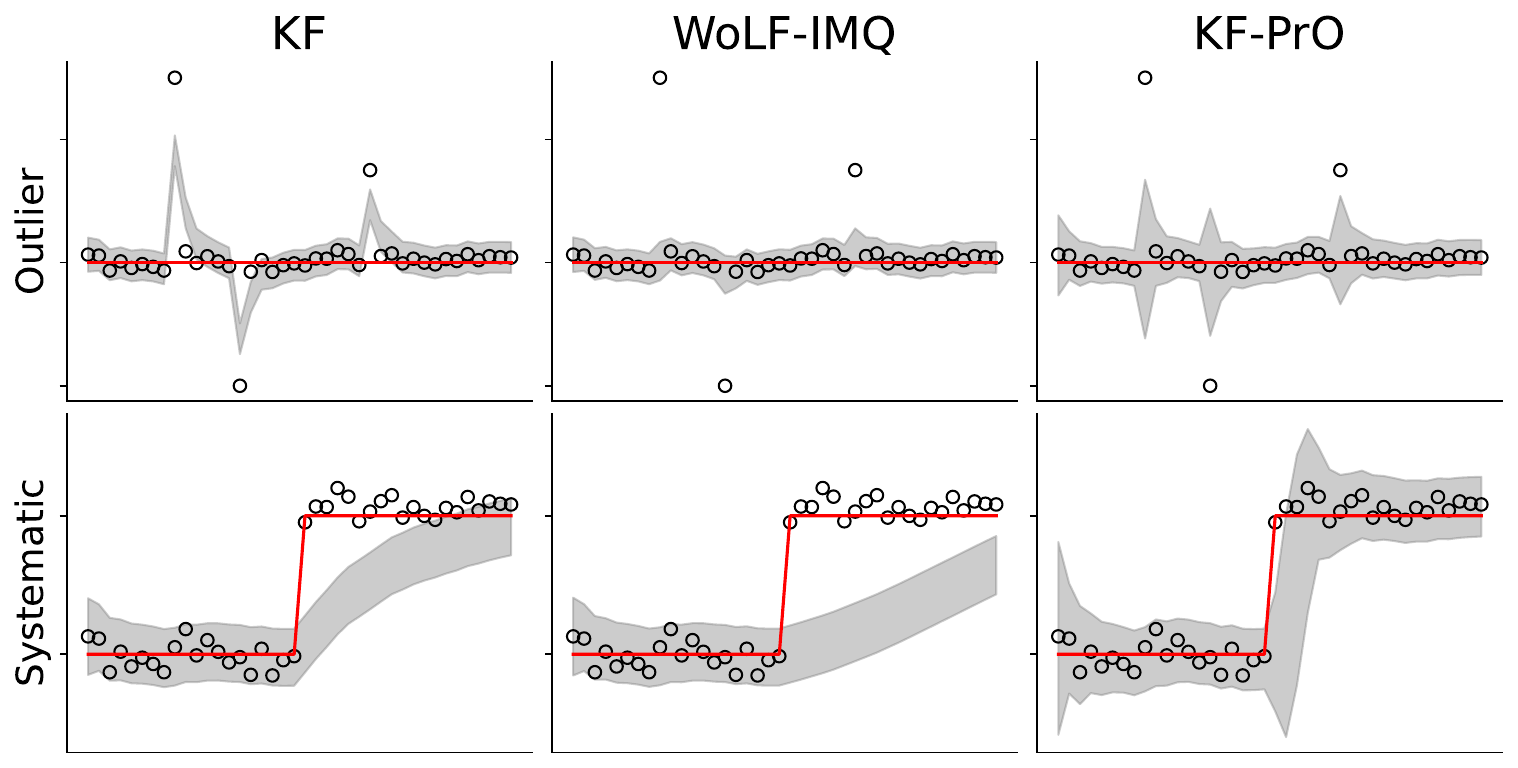}
\caption{Illustration on the Gaussian location model. 
Data $(y_k)_{k=1}^n$ are shown as circles and the true latent state is in red; in the top row the data are corrupted by outliers, while in the bottom row the model is systematically misspecified.
Posterior predictives 
are displayed for \mkf (left), a generalised Bayes method called \mWlfImq due to \citet{duran2024outlier} (centre) and our proposed \mPrOKF (right).
}
\label{fig: example}
\end{figure*}

\Cref{subsec: Op-cen filter} casts probabilistic filtering as an optimisation task, relating common filtering algorithms to optimisation objectives that are implicitly being minimised. 
\Cref{subsec: Pro GF} builds on this perspective to introduce a novel \ac{PrO} filtering algorithm in terms of its associated optimisation objective, initially for the case where $f$ and $h$ are linear, and \Cref{sec: kalman compare} compares the \ac{PrO} filtering distribution with standard \mkf.
The general case of the \ac{PrO} Kalman filter, where $f$ and $h$ can be nonlinear, is addressed by the \mPrO algorithm, presented in \Cref{subsec: PrO EKF}.

\subsection{Filtering as an Optimisation Task}
\label{subsec: Op-cen filter}

This section casts probabilistic filtering as an optimisation task.
The latent state $x_k$ is assumed to take values in $\mathbb{R}^d$ and we let $\mathcal{P}(\mathbb{R}^d)$ denote the set of probability distributions on $\mathbb{R}^d$.
The conditional distribution of $x_k$ (resp. $y_k$) given $x_{k-1}$ implied by \eqref{eq: SS1} and \eqref{eq: SS2} is denoted $p(x_k | x_{k-1})$ (resp. $p(y_k | x_k)$), and a prediction for $x_k$ given the observations $y_{1:l}$ is denoted $q_{k|l} \in \mathcal{P}(\mathbb{R}^d)$, where to improve presentation we interchange freely between densities and distributions, and use the standard filtering notation in \citet{Sarkka23}.

Through the lens of optimisation, existing Bayesian and generalised Bayesian filtering methods can be formulated as minimising objective functionals defined on $\mathcal{P}(\mathbb{R}^d)$.
Following standard filtering terminology, we present this in two steps: the \emph{prediction} step uses the filtering distribution from the previous step $q_{k-1|k-1}$ to calculate the latent predictive distribution $q_{k|k-1}$, then the \emph{update} step minimises an objective function in which $q_{k|k-1}$ appears as a regulariser via the \ac{KLD}:

\paragraph{Prediction Step} 

Given the filtering distribution from the previous time step, $q_{k-1|k-1}(x_{k-1})$, the latent predictive distribution is obtained using the Chapman--Kolmogorov equation
\begin{align}
	q_{k|k-1}(x_k) := \int p(x_k | x_{k-1}) q_{k-1|k-1}(x_{k-1})\mathrm{d} x_{k-1}.  \label{eq: CK}
\end{align}

\paragraph{Update Step} 

Given the latent predictive distribution in \eqref{eq: CK}, we can then formulate the filtering distribution as the solution of an infinite-dimensional optimisation problem on the set of probability measures 
\begin{align}
	q_{k|k} := \argmin_{q \in \mathcal{P}(\mathbb{R}^d)} \quad \ell (y_k; q) + \mathrm{KL}(q \Vert q_{k|k-1}) \label{eq: prob filter}
\end{align}
where the loss function $\ell$ is to be specified.

\noindent
This optimisation-centric formulation unifies Bayesian and generalised Bayesian filtering,  which correspond to different choices of loss function $\ell$ in \eqref{eq: prob filter}:

\begin{example}[Bayes; \mkf]
    Standard Bayesian inference is recovered using the (negative) average log-likelihood loss  
    \begin{align}
    \ell^{\mathrm{(Bayes)}}(y_k; q):=-\int \log (p(y_k | x_k)) q(x_k)\mathrm{d}x_k , \label{eq: Bayes loss}
    \end{align}
    a result which follows from the Donsker--Varadhan characterisation of the \ac{KLD} \citep[see e.g.][]{knoblauch2019generalized}.
    For linear $f$, $h$ and Gaussian $\epsilon_k$, $\xi_k$, the above updates exactly implement the \emph{\mkf}.
    More broadly, this perspective encompasses methods which attempt Bayesian inference for a variety of more flexible state-space models, mentioned in \Cref{sec: intro}.
\end{example}

\begin{example}[Generalised Bayes; WoLF]
    As an exemplar of generalised Bayesian filtering methods, the WoLF algorithm of \citet{duran2024outlier} employs
    $$
    \ell^{\mathrm{(WoLF)}}(y_k; q) :=-\int w(y_k, \hat{y}_k) \log (p(y_k | x_k)) q(x_k)\mathrm{d}x_k
    $$
    where $w$ is a non-negative \emph{weight function}, comparing a prediction $\hat{y}_k$ based on $y_{1:k-1}$ to the actual datum $y_k$ observed.
    The weight function is user-specified; for example as an inverse multi-quadric weight function gives rise to the \emph{\mWlfImq} method, while a variant based on Mahalanobis distance was termed \emph{\mWlfMd}.
    In either case, if the observed $y_k$ is far from the predicted $\hat{y}_k$, then $y_k$ is deemed to be unreliable and its influence on the filtering distribution is reduced.
\end{example}

The principal drawback of (generalised) Bayesian methods in this context is that they lack awareness that the state-space model could be misspecified.
The issue is seen most easily in the simplest setting where $x_k = \theta$ is a static parameter and we model $y_k = \theta + \xi_k$ as a noisy observation of $\theta$; given sufficient data, both Bayesian and generalised Bayesian posteriors collapse onto a single ``best'' parameter $\theta_\star$ \citep{miller2021asymptotic}, while such arbitrarily confident predictions are inappropriate when the state-space model on which they are based is misspecified.  
To address this problem we adopt an alternative, predictively-oriented perspective next.

\subsection{Predictively-Oriented Kalman Filtering}
\label{subsec: Pro GF}

Our novel \ac{PrO} filtering method employs a loss function which explicitly promotes \emph{predictive} performance of the filtering distribution.
Taking inspiration from the earlier work of \citet{masegosa2020learning,jankowiak2020parametric,sheth2020pseudo,morningstar2022pacm,lai2024predictive,shenprediction,mclatchie2025predictively,liu2025detecting} in the static inference context, we define the \ac{PrO} filtering objective as follows:

\begin{definition}[\ac{PrO} filtering objective]
    The \emph{\ac{PrO} filter} employs the loss 
    \begin{align}
    \ell^{(\mathrm{PrO})}(y_k; q) :=-\log \left(\int p(y_k | x_k) q(x_k)\mathrm{d}x_k\right) . \label{eq: pro loss}
    \end{align}
\end{definition}

Unlike attempts to perform Bayesian inference with a more sophisticated 
state-space model, or generalised Bayesian solutions in which additional hyper-parameters must be user-specified, the \ac{PrO} filter performs a probabilistic \textit{lifting} of the original measurement model
$$
\begin{tikzpicture}[baseline=(A.base)]
\node (A) {$p(y_k | x_k) $};
\node (B) [right=1cm of A] {$p_q(y_k) := \int p(y_k | x_k) q(x_k)\mathrm{d}x_k$};
\draw[->, looseness=1, out=20, in=160] (A) to node[above] {} (B);
\end{tikzpicture}
$$
viewing $q$ (rather than $x_k$) as the `parameter' of the model, in this case an (infinite) mixture model in a similar manner to nonparametric maximum likelihood \citep{laird1978nonparametric}.
As such, our construction circumvents the requirement to manually design a more sophisticated model, and does not introduce any hyper-parameter which need to be specified (cf. \Cref{tab:complexity-linear-model}).

Compared to the Bayesian loss \eqref{eq: Bayes loss}, the order of the logarithm and the integral are interchanged; this key difference causes the \ac{PrO} filter to behave in a more desirable way compared to (generalised) Bayesian methods when the state-space model is misspecified. 
Indeed, considering $y_k$ to be a sample from the true data-generating distribution $p_{\text{true}}$, we can see that
\begin{align}
	\mathbb{E}_{y_k\sim p_{\text{true}}(\cdot | x_k)}[\ell^{(\text{PrO})}(y_k ; q)] & = \underbrace{\mathrm{KL}\left(p_{\text{true}}(\cdot | x_k) \left\Vert p_q \right.\right)}_{\text{predictive fit}} \quad + \; \mathrm{constant} , \label{eq: min KL}
\end{align}
which demonstrates that \ac{PrO} filtering explicitly minimises the \ac{KLD} between the predictive $p_q$ and the true data distribution $p_{\text{true}}(\cdot | x_k)$, automatically calibrating predictions even when the model is misspecified \citep{mclatchie2025predictively}.

\begin{remark}[General scoring rules]
Alternative scoring rules could be used in \eqref{eq: pro loss} and would correspond to alternative divergences in \eqref{eq: min KL} \citep{gneiting2007strictly}.
However, the log scoring rule ensures that both the loss and the regulariser in \eqref{eq: prob filter} are in the same units (i.e. nats), avoiding the need to introduce a `learning rate' to balance the size of the two terms \citep[][Remark 2]{liu2025detecting}.     
\end{remark}

The key technical question that we address in this work is how to efficiently (and perhaps approximately) perform the update in \eqref{eq: prob filter} when the \ac{PrO} filtering objective is used.
Techniques such as gradient flows and mean-field Langevin dynamics can in principle be applied to approximate $q_{k|k}$ \citep{shenprediction,mclatchie2025predictively,chazal2025computable}, but they are computationally intensive and therefore impractical when rapid calculation is required, as is often the case in a filtering context.
Instead, we develop a fast update analogous to the \mkfExtended, based on recursive Gaussian variational approximations~\citep{lambert2022recursive, jones2024bayesian}, which we call the \mPrO.

\paragraph{Computing the PrO Filter}

For presentational purposes we begin by considering the simplest case where $f$, $h$ are linear and $\epsilon_k$, $\xi_k$ are Gaussian, i.e. the setting of the \mkf:
\begin{align*}
    x_k | x_{k-1} & \sim \mathcal{N}(A_{k-1}x_{k-1}, \Lambda_{k-1}) \\
	y_k | x_k & \sim \mathcal{N}(H_k x_k, R_k) 
\end{align*}
for positive definite $\Lambda_{k-1}$ and $R_k$.
The optimal filtering distribution $q$ minimising the \ac{PrO} objective~\eqref{eq: prob filter} is not necessarily Gaussian, even with a Gaussian prior $q_{k|k-1}$. 
To relieve the computational burden, we define the \mPrOKF as the optimal filtering distribution within the Gaussian variational family
\begin{align}
	q_{k|k} := \argmin_{q \in \mathcal{Q}(\mathbb{R}^d) }\; -\log \left(\int p(y_k | x_k) q(x_k)\mathrm{d}x_k\right) + \mathrm{KL}(q \Vert q_{k|k-1}),  \label{eq: pro opt task}
\end{align}
where $\mathcal{Q}(\mathbb{R}^d)$ is the set of all Gaussian distributions on $\mathbb{R}^d$. 

Given a filtering distribution $q_{k-1|k-1}$ from the previous time step of the form $\mathcal{N}(m_{k-1}, P_{k-1})$, the \textbf{prediction step} coincides with the standard \mkf, as the Chapman--Kolmogorov equation yields a closed-form solution 
\begin{align}
    q_{k|k-1}=\mathcal{N}(\tilde{m}_k, \tilde{P}_k), \qquad 
	\tilde{m}_k = A_{k-1} m_{k-1}, \qquad \tilde{P}_k = A_{k-1}P_{k-1}A_{k-1}^\top + \Lambda_{k-1}.  \label{eq: predictive step}
\end{align}
However, the \textbf{update step} does not have a closed form in general.
Nevertheless, the optimisation objective in \eqref{eq: pro opt task} can be explicitly computed:

\begin{proposition}[Explicit form of PrO filtering objective]
\label{prop: explicit J}
Parametrising $q \in \mathcal{Q}(\mathbb{R}^d)$ as $q = \mathcal{N}(m, P)$, we can write down the objective function in \eqref{eq: pro opt task} as 
\begin{align}
	 \mathcal{J}_k(m, P)\: \stackrel{+C}{:=}&\; \frac{1}{2}\bigg[\mathrm{tr}\left(\tilde{P}_k^{-1} P\right)+\left\Vert \tilde{P}_k^{-\frac{1}{2}} \left(m - \tilde{m}_k\right)\right\Vert^2 + \log \frac{|S_k(P)|}{|P|} + \left\Vert S_k^{-\frac{1}{2}}(P) v_k(m)\right\Vert^2\bigg] \label{eq: pro_gaussian}
\end{align}    
where $S_k(P):= H_k PH_k^\top + R_k$ and $v_k(m):= y_k-H_k m$.
\end{proposition}

The proof of \Cref{prop: explicit J} is contained in \Cref{app: explicit J}.
To minimise \eqref{eq: pro_gaussian}, we first note that $\mathcal{J}_k$ is convex in $m$ when $P$ is fixed, and that the minimiser $m_k^\star(P)$ can be explicitly computed:

\begin{proposition}[Explicit minimisation over $m$ for fixed $P$]
\label{prop: mini over m}
Let $\mathcal{J}_k$ be the objective function in \eqref{eq: pro_gaussian} and let the covariance matrix $P$ be fixed.
Then $m \mapsto \mathcal{J}_k(m,P)$ is uniquely minimised at
\begin{align}
    m_k^\star(P) = \tilde{m}_k + K_k(P) (y_k - H_k \tilde{m}_k), \label{eq: mean update}
\end{align}
where $K_k(P) := [ \tilde{P}_k^{-1} + H_k^\top S_k^{-1}(P) H_k ]^{-1} H_k^\top S_k(P)^{-1}$.
\end{proposition}

The proof of \Cref{prop: mini over m} is contained in \Cref{app: mini over m}.
Similarly to the \mkf, \eqref{eq: mean update} takes the form of a prior mean ($\tilde{m}_k$) plus a `correction' term which depends on how much the observed datum ($y_k$) deviates from the predictive mean ($H_k \tilde{m}_k$).
The matrix $K_k$ in \eqref{eq: mean update} is an analogue of the \emph{Kalman gain} matrix for the \mPrOKF; we elaborate on this relationship in \Cref{sec: kalman compare}.

Although $\Phi_k(P) := \mathcal{J}_k(m_k^\star(P), P)$ is non-convex in $P$, we note that $- \log |S_k(P)|$ is convex in $P$, and the remainder is convex, forming a \ac{DoC} structure~\citep{tao1997convex,yuille2003concave}: 

\begin{proposition}[Difference of convex representation for the PrO filtering objective]
\label{prop: DoC}
\begin{align*}
\Phi_k(P)
\stackrel{+C}{=}
\frac{1}{2}\Big[
\underbrace{ \mathrm{tr}(\tilde{P}_k^{-1} P)
- \log |P|
+ v_k(\tilde{m}_k)^\top \big(S_k(P) + H_k \tilde{P}_k H_k^\top\big)^{-1} v_k(\tilde{m}_k) }_{\text{\normalfont convex in $P$}}
- \underbrace{ - \log |S_k(P)| }_{\text{\normalfont convex in $P$}}
\Big].
\end{align*}
\end{proposition}

The proof of \Cref{prop: DoC} is contained in \Cref{app: DoC}.
Due to the \ac{DoC} structure, efficient numerical methods from \ac{DoC} programming can be applied to obtain the \mPrOKF covariance $P_k = \argmin_P \Phi_k(P)$, and hence also the \ac{PrO} filtering mean $m_k = m_k^\star(P_k)$; full implementational details are contained in \Cref{app: DoC methods}.

\subsection{Comparison with the Standard Kalman Filter}
\label{sec: kalman compare}

This section compares the one-step behaviour of \mkf with our proposed \mPrOKF, with the main findings summarised in the following result:

\begin{proposition}[Conservativity of PrO filtering relative to \mkf]
\label{prop: comparison}
Let $\KFmean$ and $\KFcov$ denote the mean and covariance associated with the \emph{\mkf}, and assume both the \emph{\mkf} and \emph{\mPrOKF} start from a common prior $q_{k-1|k-1}$, with predictive mean $\tilde{m}_k$ defined in \eqref{eq: predictive step}.
Then
\begin{enumerate}
    \item $\lVert m_k - \tilde{m}_k\rVert \leq \lVert \KFmean - \tilde{m}_k\rVert$
    \item $\nabla \Phi_k(\KFcov) \preceq 0$.
\end{enumerate}
\end{proposition}

The proof of \Cref{prop: comparison} is contained in \Cref{app: comparison}.
The first part shows that the \mPrOKF mean $m_k$ is no more sensitive to $y_k$ than the \mkf, while the second part shows that inflating the covariance (i.e. $\KFcov \mapsto \KFcov + \Delta \KFcov$ for a positive semi-definite perturbation $\Delta \KFcov$) is a local descent direction of $\Phi_k$ at the Kalman solution $\KFcov$, indicating that minimising $\Phi_k$ locally promotes higher posterior uncertainty compared to the \mkf.
(In dimension $d=1$ the function $\Phi$ is convex, implying that the \mPrOKF variance $P_k$ is at least as large as the \mkf variance $\KFcov$.)
Further, an outlier in $y_k$ results in a steeper gradient $\nabla \Phi_k(\KFcov)$, leading to \emph{data-dependent covariance inflation} relative to the \mkf\footnote{The posterior covariance $\KFcov$ of \mkf is $y_k$-independent, a consequence of assuming the state-space model is well-specified.} when outliers are encountered.

\Cref{fig: example} illustrates the difference in behaviour of the \mkf and \mPrOKF in the setting of a simple state space model: $A_k=1, R_k=\sigma^2=1, \Lambda_k=\epsilon^2=10^{-4}, H_k=1$. 
The dataset in the first row is corrupted by outliers, and we observe both the conservative mean update and the data-driven covariance inflation for \mPrOKF. 
The second row represents a missepcified dynamics model, where the true state abruptly changes after $k = 20$. 
Like the \mkf, the \mPrOKF mean $m_{21}$ is not strongly affected by the first outlier $y_{21}$ (cf. Part 1 of \Cref{prop: comparison}), however unlike the \mkf the \mPrOKF covariance $P_{21}$ is increased (cf. Part 2 of \Cref{prop: comparison}).
This large $P_{21}$ enables the \mPrOKF to rapidly update the filter mean $m_{22}$ at the next step using~\eqref{eq: mean update}, resulting in better predictions compared with \mkf and \mWlfImq for all subsequent values of $k$ considered.

\subsection{A PrO Extended Kalman Filter}
\label{subsec: PrO EKF}

The \mkfExtended extends the \mkf using local linearisation to perform approximate inference in settings where $f$ and $h$ are nonlinear \citep{maybeck1982stochastic}. 
Analogously, in this section we extend the \mPrOKF to the \mPrO. 
Firstly, the \textbf{prediction step} follows the same recipe as the \mkfExtended \citep[][Algorithm 7.5]{Sarkka23}, linearising $f$ around the mean from the previous time step (i.e. $m_{k-1}$):
\begin{align*}
    \tilde{m}_k &= f(m_{k-1}, 0), \\
    \tilde{P}_k &= \mathbf{F}_x(m_{k-1}, 0)\: P_{k-1}\: \mathbf{F}_x(m_{k-1}, 0)^\top + \mathbf{F}_\epsilon(m_{k-1}, 0)\: \Lambda_{k-1}\: \mathbf{F}_\epsilon(m_{k-1}, 0)^\top ,
\end{align*}
where without loss of generality we assume that $\epsilon_k \sim \mathcal{N}(0, \Lambda_k)$ and $\xi_k \sim \mathcal{N}(0, R_k)$.
Here $\mathbf{F}_x$ and $\mathbf{F}_\epsilon$ denote the Jacobian matrices of the dynamic model $f$ in \eqref{eq: SS1}. 
For the \textbf{update step}, we could proceed similarly to the \mkfExtended and linearise $h$ around the predictive mean from the previous time step (i.e. $\tilde{m}_k$).
However the \ac{DoC} formulation from \Cref{subsec: Pro GF} affords an opportunity to linearise around a more suitable origin, namely $m_k^\star = m_k^\star(P)$ where $P$ is the current iterate on the \ac{DoC} optimisation path, analogous to the \emph{iterative} \mkfExtended (see \citealp{gelb1974applied}, and Algorithm 7.9 in \citealp{Sarkka23}).
Thus we take
\begin{align}
    S_k(P) &= \mathbf{H}_{x}(m_k^\star, 0)\: P\:\mathbf{H}_{x}(m_k^\star, 0)^\top + \mathbf{H}_\xi(m_k^\star, 0)\: R_k\: \mathbf{H}_\xi(m_k^\star, 0)^\top, \label{eq: pro linear1}\\
    v_k(m) &= y_k - h(m_k^\star, 0) - \mathbf{H}_m(m_k^\star, 0)(m-m_k^\star) \label{eq: pro linear2},
\end{align}
where $\mathbf{H}_x, \mathbf{H}_\xi$ refer to the Jacobian matrices of the measurement model $h$ in \eqref{eq: SS2}.

\section{Empirical Results}
\label{sec: experiments}

This section presents a range of results to demonstrate the properties of the \mPrO, relative to baselines described in \Cref{sec: baselines}.
Tracking of a 2D object is considered in \Cref{experiment:2d-tracking}, and prediction for a chaotic dynamical system is considered in \Cref{subsec: Lorenz}.

\subsection{Baseline Methods}
\label{sec: baselines}

For baselines, we consider the standard EKF (\mkfExtended), the iterated EKF (\mkfExtendedIterated; \citealp{gelb1974applied}) and also methods that are representative of recent state-of-the-art approaches to robust filtering:
the Bernoulli filter of \citet{wang2018} (\mWangExtended; an example of a \emph{detect-and-reject} strategy, identifying outliers and then ignoring them);
the inverse-Wishart filter of \citet{Agamennoni2012} (\mAgamenoniExtended; an example of a \emph{compensation-based} strategy, which anticipates heavier-than-Gaussian tails in the dataset), 
and the weighted observation likelihood filter \citet{duran2024outlier} (a generalised Bayesian approach, here in two variants, \mWlfImq and \mWlfMd).
Note that we do not compare against more sophisticated hierarchical methods nor methods based on particle filtering, because typically these do not scale well to high-dimensional state spaces; the computational complexity of \mPrO is designed to be comparable to the fast baseline methods, as shown in \Cref{tab:complexity-linear-model}.

\begin{table}[t!]
    \small
    \centering
    \begin{tabular}{llll}
    \toprule
               Method & Cost & \#HP & Ref \\
    \midrule
         \mkfExtended &  $O(d^3)$ & 0 & \citet{kalman1960}\\
         \mkfExtendedIterated &  $O(I\, d^3)$ & 0 & \citet{jazwinski1970stochastic}  \\
         \mWangExtended & $O(I\,d^3)$ & 2 & \citet{wang2018}\\
         \mAgamenoniExtended & $O(I\,d^3)$ & 1 & \citet{Agamennoni2012} \\
         \mWlfImq  &  $O(d^3)$ & 1 & \citet{duran2024outlier}\\
         \mWlfMd  &  $O(d^3)$ & 1 & \citet{duran2024outlier}\\
         \mPrO & $O(I\,d^3)$ & 0 & (Ours) \\
    \bottomrule
    \end{tabular}
    \caption{
        Computational complexity of the update step for each baseline method.
        Here $d$ is the dimension of the latent state, $I$ is the number of inner-loop iterations that are performed, and \#HP is the number of hyper-parameters to be specified.
    }
    \label{tab:complexity-linear-model}
\end{table}

\subsection{Tracking an Object in 2D}
\label{experiment:2d-tracking}

\begin{figure}
    \centering
    \includegraphics[width=0.9\linewidth]{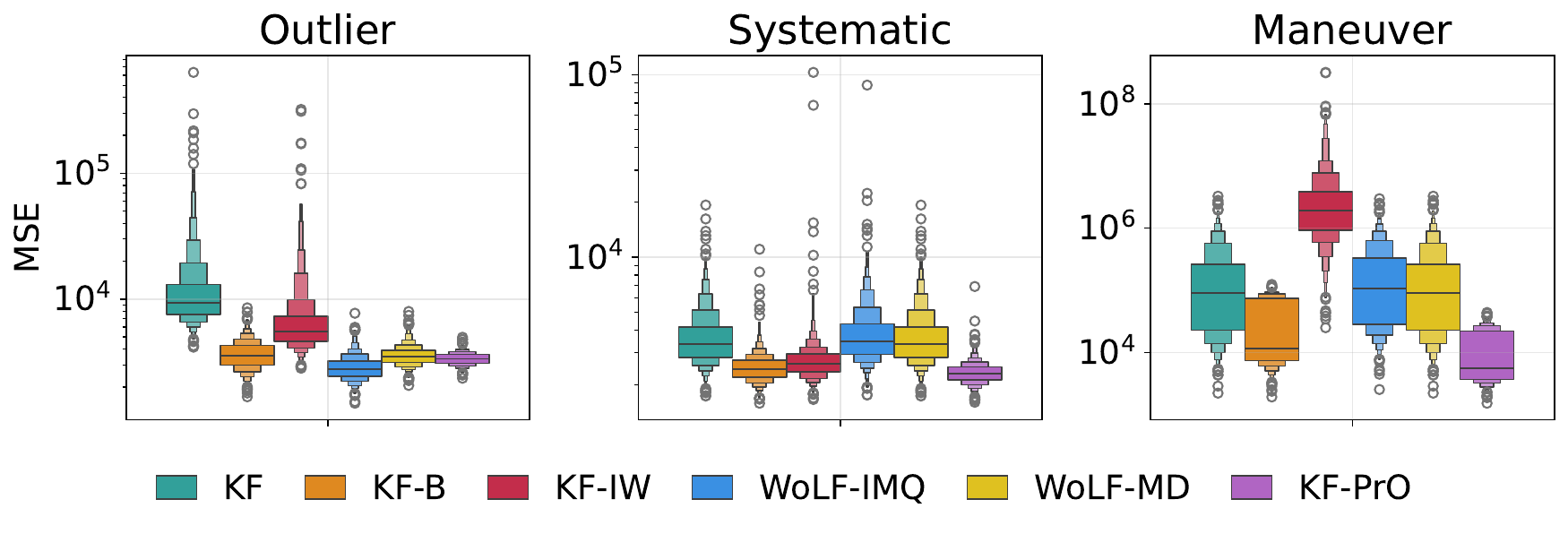}
    \caption{Tracking an object in 2D. Here the mean square error for the filtering mean of the object location is displayed for each of the methods in \Cref{tab:complexity-linear-model}, with either \textbf{outlier}, \textbf{systematic}, or \textbf{maneuver}-type misspecification present in the dataset; cf. \Cref{experiment:2d-tracking}.
    Note that in this linear setting both \mkfExtended and \mkfExtendedIterated are identical to the classical \mkf.
    A total of 500 replicates were performed.
    }
    \label{fig:2d tracking}
\end{figure}

For our first experiment we consider the classical problem of tracking an object moving with constant velocity in 2D (see e.g., Example 8.2.1.1 in \citealp{pml2Book} or Example 4.5 in \citealp{Sarkka23}).
A typical linear state-space model for this task is $f(x,\epsilon) = A + \Lambda^{1/2} \epsilon$ and $h(x,\xi) = H x + R^{1/2} \xi$, where
\begin{gather*}
    A =
    \begin{pmatrix}
    I_2 & 0.1 I_2 \\
    \mathbf{0} & I_2
    \end{pmatrix}, \quad  
    H = \begin{pmatrix}
        I_2 & \mathbf{0}
    \end{pmatrix} , \quad \Lambda = 0.1 I_4, \quad R = 10 I_2
\end{gather*}
with the state variable $x_k \in \mathbb{R}^4$ representing the location and velocity of the object.
Three types of data are considered, none of which are accurately described by the model; \textbf{systematic} misspecification draws $\epsilon_k \sim t(3)$, so that the state noise is sampled from a student-$t$ distribution with $3$ degrees of freedom, \textbf{outlier} misspecification draws $\xi_k \sim t(2.01)$, so that the measurement noise is heavier-tailed, and \textbf{maneuver} introduces random piecewise constant shifts to the velocity component, with full experimental protocol in \Cref{app: tracking}.
For this experiment we simulated $T=1,000$ time steps in total.

\Cref{fig:2d tracking} shows that, when only the measurement model is misspecified (\textbf{outlier}), the data-driven approach to the mean update enables \mWlfImq to achieve the lowest positional \ac{MSE}; however, when misspecification occurs in the dynamical model (\textbf{systematic}, and especially \textbf{maneuver}), \mPrOKF is to be preferred.

\subsection{Lorenz96 Process}
\label{subsec: Lorenz}

\begin{figure}
    \centering
    \includegraphics[width=0.8\linewidth]{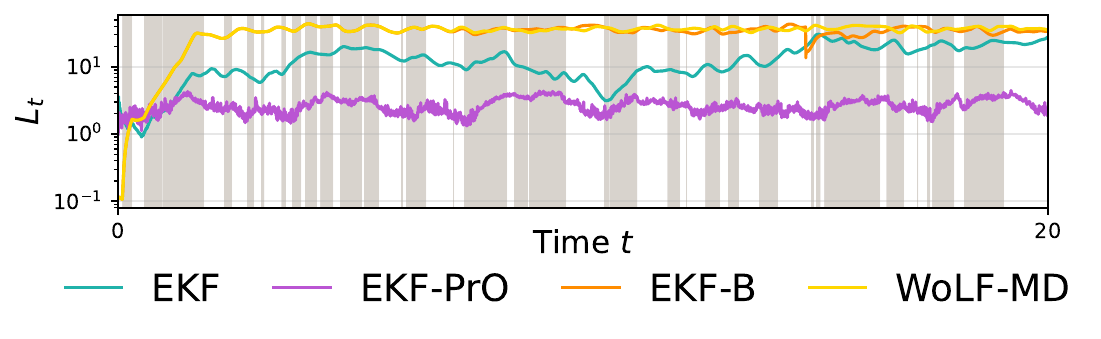}
    \caption{
    The misspecified Lorenz96 model. For a filtering algorithms with mean $\{m_t\}_{t=1}^T$, we plot the error $L_t = \lVert \theta_t - m_t\rVert$ as a function of time, and mark the region when the underlying data-generating distribution is misspecified (white means no misspecification). 
    }
    \label{fig:lorenz}
\end{figure}

For our final experiment, we consider the classical nonlinear Lorenz96 model, commonly used to simulate the atmosphere \citep{lorenz96predictability}.
We set the latent state as $d$-dimensional, $d=60$.
Our experimental protocol is similar to \citet{roth2017enekf}; full details are contained in \Cref{app: Lorenz}. The state-space model is defined by a discretised ODE system for $\theta\in\mathbb{R}^d$: $\frac{\mathrm{d}\theta_i}{\mathrm{d}t} = (\theta_{i+1}-\theta_{i-2})\theta_{i-1} - \theta_i + \phi_i$, $i \in 1, 2, \hdots, d$, and the observation $y_t$ is generated using a Gaussian likelihood. As Lorenz96 is nonlinear, we opt for the iterative variant of \mPrO. Similar to the \textbf{maneuver} example, the data-generating distribution contains an additional latent state $m_t \in \{1, 2\}$ and vectors $\phi^{(1)}, \phi^{(2)} \in \mathbb{R}^d$, such that for each $\theta_{t + \Delta}$, it takes an Euler step in the Lorenz96 system driven by $\phi^{(m_t)}$. 

Interestingly, \Cref{fig:lorenz} illustrates that outlier-robust filters alone cannot alleviate systematic misspecification, and can sometimes worsen it: \mWangExtended and \mWlfMd underperform \mkfExtended because an ``outlier'' observation can be rejected if it is too far from the predicted value from the prior, however, when the outlier is caused by state-space misspecification, this leads to consecutive rejections and slower updates in the state-space. Therefore, \mPrO is particularly beneficial in cases where the precise source of model misspecification is undetermined.

\section{Discussion}
\label{sec: discuss}

In applications it is rare for a state-space model to be well-specified, and even if this is the case at the outset, dynamics may change over time or data may become corrupted.
Existing solutions either assume the dynamical model is correct and discount observations to hedge against outliers, or include additional degrees of freedom into the model.
However, discounting observations reduces the amount of information available to the filtering method, while increasing model capacity introduces hyper-parameters which need to be elicited or estimated.
Our contribution established \mPrO,
a novel hyper-parameter-free post-Bayesian filtering method whose predictions are explicitly calibrated against the data-generating process, cf. \eqref{eq: min KL}, suitable for situations where \emph{both} the dynamical model and measurement model may be misspecified.

Our work contributes to an emerging literature on \ac{PrO} posteriors \citep{lai2024predictive,shenprediction}, extending their applicability to state-space models.
On the other hand, the theoretical properties of the \mPrO (e.g. statistical efficiency, dimension dependence) are not well-understood, and further work will be needed to extend existing theoretical analysis of \ac{PrO} posteriors to the state-space setting \citep{mclatchie2025predictively}.
Further, while all experiments ran in a few minutes on a standard laptop %
calculations for \mPrO were $10$-$20\times$ slower than the extremely rapid \mkf; reducing this wall-clock time is left as further work.

The main technical challenge was to develop an efficient (approximate) update for the filtering mean and covariance, which we achieved here with linearisation, Gaussian variational approximation, and \ac{DoC} programming.
Several extensions and improvements can be considered, based on relaxing the linearity and Gaussian assumptions (albeit potentially introducing hyper-parameters and increasing computational cost).
The \textbf{prediction step} relied on linearisation to integrate $f$ with respect to a Gaussian, but more sophisticated techniques such as moment matching~\citep{maybeck1982stochastic}, cubature~\citep{ito2002gaussian}, and unscented transforms~\citep{julier1996general} could be used.
The \textbf{update step} relied on a Gaussian approximation to facilitate optimisation, but particle methods analogous to the ensemble Kalman filter \citep{evensen2003ensemble} could also be considered.

\paragraph{Acknowledgments}
CJO and ZS were supported by EPSRC (EP/W019590/1). 
CJO was supported by a Philip Leverhulme Prize (PLP-2023-004).

\bibliographystyle{abbrvnat}
\bibliography{bibliography}

\appendix
\section*{Supplementary Material}

This document contains supplementary material for the paper \emph{Predictively-Oriented Kalman Filtering}.
\Cref{app: proofs} contains proofs for all theoretical results in the main text.
\Cref{app: DoC methods} explains how \ac{DoC} programming was implemented for the \textbf{update step} of the \mPrO.
\Cref{app: protocol} contains full details required to reproduce the empirical results in the main text.

\section{Proofs}
\label{app: proofs}

This appendix contains proofs for all theoretical results in the main text.
The proof of \Cref{prop: explicit J} is contained in \Cref{app: explicit J}, the proof of \Cref{prop: mini over m} is contained in \Cref{app: mini over m}, the proof of \Cref{prop: DoC} is contained in \Cref{app: DoC}, and the proof of \Cref{prop: comparison} is contained in \Cref{app: comparison}.

\medskip

\begin{remark}[Existence of matrix inverses]
\label{rem: inversion}
    Our standing assumption in \Cref{subsec: Pro GF} that $\Lambda_{k-1}$ and $R_k$ are (symmetric) positive definite ensures that the matrices $\tilde{P}_k$ and $S_k(P)$ exist and are also (symmetric) positive definite and can hence be inverted.
\end{remark}

\subsection{Proof of \Cref{prop: explicit J}}
\label{app: explicit J}

Denote the density function of a Gaussian with mean $\mu$ and covariance $\Sigma$ as $\mathcal{N}(\cdot ; \mu,\Sigma)$.
Parametrising $q \in \mathcal{Q}(\mathbb{R}^d)$ as $q = \mathcal{N}(m, P)$, we can write down the objective function in \eqref{eq: pro opt task} as
the convolution of two Gaussian densities:
\begin{align*}
    \mathcal{J}_k(m, P) &= -\log \int \mathcal{N}(y_k ; H_k x,  R_k)\mathcal{N}(x; m, P)\mathrm{d} x + \mathrm{KL}( \mathcal{N}(m, P)\Vert \mathcal{N}(\tilde{m}_k, \tilde{P}_k) )\\
    &= -\log \mathcal{N}( 0 ; \underbrace{ y_k - H_k m }_{= v_k(m)} , \underbrace{ H_k P H_k^\top + R_k }_{=S_k(P)} ) + \mathrm{KL}( \mathcal{N}(m, P)\Vert \mathcal{N}(\tilde{m}_k, \tilde{P}_k) )\\
    & = \frac{1}{2} \left[ \log |S_k(P)| + \left\Vert S_k^{-\frac{1}{2}}(P) v_k(m)\right\Vert^2 \right] \\
    & \qquad \qquad + \frac{1}{2} \left[ \mathrm{tr}\left(\tilde{P}_k^{-1} P\right)+\left\Vert \tilde{P}_k^{-\frac{1}{2}} \left(m - \tilde{m}_k \right)\right\Vert^2  + \log \frac{|\tilde{P}_k|}{|P|} \right] \\
    & \hspace{-10pt} \stackrel{+C}{=}  \frac{1}{2}\bigg[ \mathrm{tr}\left(\tilde{P}_k^{-1} P\right)+\left\Vert \tilde{P}_k^{-\frac{1}{2}} \left(m - \tilde{m}_k \right)\right\Vert^2 -\log |P| + \log |S_k(P)| + \left\Vert S_k^{-\frac{1}{2}}(P) v_k(m)\right\Vert^2\bigg] ,
\end{align*}
recovering the final expression in \Cref{prop: explicit J}.

\subsection{Proof of \Cref{prop: mini over m}}
\label{app: mini over m}

For fixed $P$, the map $m \mapsto \mathcal{J}_k(m, P)$ is a quadratic function; cf. \eqref{eq: pro_gaussian}. 
Therefore, again with fixed $P$, we can find the critical point using the first-order optimality condition:
\begin{align*}
    \nabla_m \mathcal{J}_k(m, P) = 0 \; \implies \; \tilde{P}_k^{-1}(m-\tilde{m}_k) - H_k^\top S_k^{-1}(P)(y_k-H_k m) = 0
\end{align*}
Rearranging the above formula,
\begin{align*}
    \left(\tilde{P}_k^{-1} + H_k^\top S_k^{-1}(P) H_k \right) m
&= \tilde{P}_k^{-1}\tilde{m}_k + H_k^\top S_k^{-1}(P) y_k \\
&= \tilde{P}_k^{-1}\tilde{m}_k + H_k^\top S_k^{-1}(P) (y_k-H_k\tilde{m}_k + H_k\tilde{m}_k)\\
&= \left(\tilde{P}_k^{-1}+H_k^\top S_k^{-1}(P) H_k \right) \tilde{m}_k
   + H_k^\top S_k^{-1}(P) (y_k-H_k \tilde{m}_k).
\end{align*}
Note that this linear system is uniquely invertible from positive-definiteness of $\tilde{P}_k^{-1}$; cf. \Cref{rem: inversion}.
Solving this linear system for $m$, we thus obtain the prediction-correction form of the unique critical point
\begin{align*}
    m_k^\star(P) = \underbrace{ \tilde{m}_k }_{\text{predict}} +  \underbrace{ \left(\tilde{P}_k^{-1}+H_k^\top S_k^{-1}(P) H_k \right)^{-1} H_k^\top S_k^{-1}(P) (y_k - H_k \tilde{m}_k ) }_{\text{correct}} ,
\end{align*}
as claimed.

\subsection{Proof of \Cref{prop: DoC}}
\label{app: DoC}

First we note that
\[
v_k(m) = y_k - H_k m = v_k(\tilde{m}_k) - H_k (m - \tilde{m}_k),
\]
so that from the definition of $\mathcal{J}_k(m,P)$ in \Cref{prop: explicit J},
\begin{align}
\mathcal{J}_k(m,P)
&\stackrel{+C}{=} \frac{1}{2}\Big[
\mathrm{tr}(\tilde{P}_k^{-1} P)
+ (m - \tilde{m}_k)^\top \tilde{P}_k^{-1} (m - \tilde{m}_k)
+ \log \frac{|S_k(P)|}{|P|} \label{eq: J expand} \\
& \qquad \qquad
+ (v_k(\tilde{m}_k) - H_k (m - \tilde{m}_k))^\top S_k^{-1}(P)
(v_k(\tilde{m}_k) - H_k (m - \tilde{m}_k))
\Big]. \nonumber
\end{align}
Let \(x := m - \tilde{m}_k\). 
The terms in \eqref{eq: J expand} depending on \(x\) are
\[
x^\top \tilde{P}_k^{-1} x
+ (v_k(\tilde{m}_k) - H_k x)^\top S_k^{-1}(P) (v_k(\tilde{m}_k) - H_k x).
\]
Expanding,
\begin{align}
&x^\top \tilde{P}_k^{-1} x
+ v_k(\tilde{m}_k)^\top S_k^{-1}(P) v_k(\tilde{m}_k)
- 2 x^\top H_k^\top S_k^{-1}(P) v_k(\tilde{m}_k)
+ x^\top H_k^\top S_k^{-1}(P) H_k x \nonumber \\
&= x^\top B x - 2 x^\top c + v_k(\tilde{m}_k)^\top S_k^{-1}(P) v_k(\tilde{m}_k), \label{eq: interm quad}
\end{align}
where $B := \tilde{P}_k^{-1} + H_k^\top S_k^{-1}(P) H_k$ and $c := H_k^\top S_k^{-1}(P) v_k(\tilde{m}_k)$.
Since \(B \succ 0\), the minimiser of \eqref{eq: interm quad} is \(x^\star = B^{-1} c\), and
\begin{align}
\min_x \left(x^\top B x - 2 x^\top c\right) = - c^\top B^{-1} c. \label{eq: min quad}
\end{align}
Recalling that $\Phi_k(P) = \min_m \mathcal{J}_k(m, P) = \min_x \mathcal{J}_k(x + \tilde{m}_k , P)$, the calculation in \eqref{eq: min quad} establishes that
\begin{align}
\Phi_k(P)
\stackrel{+C}{=}
\frac{1}{2}\Big[
\mathrm{tr}(\tilde{P}_k^{-1} P)
+ \log \frac{|S_k(P)|}{|P|}
+ v_k(\tilde{m}_k)^\top S_k^{-1}(P) v_k(\tilde{m}_k)
- c^\top B^{-1} c
\Big].  \label{eq: phi p init}
\end{align}
From the definition of $B$ and $c$,
\begin{align}
c^\top B^{-1} c
=
v_k(\tilde{m}_k)^\top S_k^{-1}(P)
H_k
\big(\tilde{P}_k^{-1} + H_k^\top S_k^{-1}(P) H_k\big)^{-1}
H_k^\top
S_k^{-1}(P)
v_k(\tilde{m}_k).  \label{eq: Phi k interm}
\end{align}
Using the Woodbury matrix inversion identity,
\[
S_k^{-1}(P)
-
S_k^{-1}(P) H_k
\big(\tilde{P}_k^{-1} + H_k^\top S_k^{-1}(P) H_k\big)^{-1}
H_k^\top S_k^{-1}(P)
=
\big(S_k(P) + H_k \tilde{P}_k H_k^\top\big)^{-1}.
\]
Combining this identity with \eqref{eq: Phi k interm},
\begin{align}
v_k(\tilde{m}_k)^\top S_k^{-1}(P) v_k(\tilde{m}_k) - c^\top B^{-1} c
=
v_k(\tilde{m}_k)^\top \big(S_k(P) + H_k \tilde{P}_k H_k^\top\big)^{-1} v_k(\tilde{m}_k).  \label{eq: final step}
\end{align}
Substituting \eqref{eq: final step} back into \eqref{eq: phi p init} gives the claimed result.

\subsection{Proof of \Cref{prop: comparison}}
\label{app: comparison}

In the statement and proof of \Cref{prop: comparison}, we use $\preceq$ to denote the Loewner ordering on the cone of (real) symmetric positive semi-definite matrices; i.e. $A \preceq B$ if and only if $B - A$ is a positive semi-definite matrix.

The standard \mkf update equations are
\begin{align}
    \KFcov^{-1} &= \tilde{P}_k^{-1} + H_k^\top R_k^{-1} H_k, \label{eq: KF P} \\ 
    \KFmean &= \tilde{m}_k + \KFcov H_k^\top R_k^{-1}(y_k -H_k \tilde{m}_k) = \tilde{m}_k + \KFgain v_k(\tilde{m}_k) . \label{eq: KF m} 
\end{align}
where $\KFgain := \KFcov H_k^\top R_k^{-1}$ is the classical Kalman gain matrix.

To establish the first claim, we wish to compare
\begin{align*}
    \KFmean - \tilde{m}_k & = \KFgain v_k(\tilde{m}_k) \\
    m_k - \tilde{m}_k & = K_k(P_k) v_k(\tilde{m}_k) .
\end{align*}
Recall that
$$
K_k(P_k) = \underbrace{ [ \tilde{P}_k^{-1} + H_k^\top S_k^{-1}(P_k) H_k ]^{-1} }_{ =: \hat{P}_k } H_k^\top S_k(P_k)^{-1} .
$$
and notice that $\hat{P}_k \preceq \tilde{P}_k$.
Then 
\begin{align}
K_k(P) K_k(P)^\top & = \hat{P}_k H_k^\top S_k^{-2}(P_k) H_k  \hat{P}_k^\top \nonumber  \\
& \preceq \tilde{P}_k H_k^\top S_k^{-2}(P_k) H_k \tilde{P}_k^\top \label{eq: tricky 1} \\
& \preceq \tilde{P}_k H_k^\top R_k^{-2} H_k \tilde{P}_k^\top  \label{eq: tricky 2} \\
& = \KFgain \KFgain^\top   \nonumber
\end{align}
where \eqref{eq: tricky 1} follows from $\hat{P}_k \preceq \tilde{P}_k$ and \eqref{eq: tricky 2} follows from $R_k \preceq H_k P_k H_k^\top + R_k = S_k(P_k)$.
Thus, 
\[
v_k(\tilde{m}_k)^\top K_k(P_k) K_k(P_k)^\top v_k(\tilde{m}_k)
\le
v_k(\tilde{m}_k)^\top \KFgain \KFgain^\top v_k(\tilde{m}_k),
\]
which implies $\|m_k - \tilde m_k\| \le \|\KFmean - \tilde m_k\|$, as claimed.

To establish the second claim, recall from \Cref{prop: DoC} that
\begin{align*}
\Phi_k(P)
\stackrel{+C}{=}
\frac{1}{2}\Big[
\mathrm{tr}(\tilde{P}_k^{-1} P)
- \log |P|
+ v_k(\tilde{m}_k)^\top \big( \underbrace{ S_k(P) + H_k \tilde{P}_k H_k^\top }_{ =: M_k(P) } \big)^{-1} v_k(\tilde{m}_k) 
+ \log |S_k(P)| 
\Big].
\end{align*}
Computing the gradient, using the standard
identities $\nabla_P\mathrm{tr}(AP)=A^\top$, $\nabla_P\log|P|=P^{-\top}$,
$\nabla_P\log|f(P)|=(\partial f/\partial P)^\top f^{-\top}$, and
$\nabla_P[v^\top f(P)^{-1}v]=-(\partial f/\partial P)^\top f^{-\top}vv^\top f^{-\top}$, we obtain that
\begin{align*}
  \nabla_P \Phi_k(P)
  =
  \frac{1}{2}\Bigl[
    \underbrace{ \tilde{P}_k^{-1}
    - P^{-1}
    + H_k^\top S_k^{-1}(P) H_k }_{(*)}
    \underbrace{ - H_k^\top M_k^{-1}(P) v_k(\tilde{m}_k) v_k(\tilde{m}_k)^\top M_k^{-1}(P) H_k }_{ \preceq 0 }
  \Bigr]
\end{align*}
where the negative semi-definiteness of the second term follows from noting this term is rank-1.
To complete the proof we will show that $(*) \preceq 0$ at $P = \KFcov$.
Indeed, at $P = \KFcov$, and using the definition of $\KFcov$ in \eqref{eq: KF P},
\begin{align*}
    (*) & = \tilde{P}_k^{-1}
    - \KFcov^{-1}
    + H_k^\top S_k^{-1}(\KFcov) H_k \\
    & = \tilde{P}_k^{-1} - [ \tilde{P}_k^{-1} + H_k^\top R_k^{-1} H_k ] + H_k^\top S_k^{-1}(\KFcov) H_k \\
    & = H_k^\top [ S_k^{-1}(\KFcov) - R_k^{-1} ] H_k \\
    & \preceq 0
\end{align*}
where the final conclusion follows from $R_k \preceq H_k \KFcov H_k^\top + R_k = S_k(\KFcov)$.

\section{Difference-of-Convex Optimisation}
\label{app: DoC methods}

As the objective function $\Phi_k(P)$ in \Cref{prop: DoC} can be written as the difference of two convex functions, it is possible to apply \ac{DoC} programming to numerically obtain the \mPrO covariance matrix $P_k = \argmin_P \Phi_k(P)$. 

\paragraph{Outline of the Algorithm}

A \ac{DoC} programme is an iterative algorithm where, given the current approximation $P_k^{(t)}$, we first linearise $P \mapsto \log |S_k(P)|$ around $P_k^{(t)}$ using its first-order Taylor expansion
\begin{align}
    \log |S_k(P)| \leq \log |S_k(P_k^{(t)})| + \mathrm{tr}(S_k^{-1}(P_k^{(t)})(S_k(P)-S_k(P_k^{(t)}))) , \label{eq: log ineq}
\end{align}
where the inequality follows from concavity of $P \mapsto \log |S_k(P)|$.
Then we can combine \Cref{prop: DoC} and \eqref{eq: log ineq} to obtain an upper bound on $\Phi_k(P)$ of the form
\begin{align*}
\phi_k(P; P_k^{(t)}) 
& \stackrel{+C}{=}
\frac{1}{2}\Big[
\underbrace{ \mathrm{tr}(\tilde{P}_k^{-1} P)
- \log |P|
+ v_k(\tilde{m}_k)^\top \big(S_k(P) + H_k \tilde{P}_k H_k^\top\big)^{-1} v_k(\tilde{m}_k) }_{\text{\normalfont convex in $P$}} \\
& \qquad \qquad + \underbrace{ \mathrm{tr}(S_k^{-1}(P_k^{(t)}) S_k(P) ) }_{\text{linear in $P$}}
\Big].
\end{align*}
Since $\phi_k(\cdot ; P_k^{(t)})$ is the sum of a convex and a linear term, it is globally convex and therefore amenable to being efficiently optimised.
Thus, in the final step we use efficient convex optimisation \citep[\texttt{L-BFGS}][]{liu1989limited} to compute $P_k^{(t+1)} = \argmin_P \phi_k(P ; P_k^{(t)})$.
The iterative algorithm proceeds in this manner until convergence, as described in \Cref{alg:lbfgs}.

\begin{algorithm}[t!]
\caption{One-step update for \mPrO}
\label{alg:lbfgs}
\begin{algorithmic}[1] %
\Require previous mean $m_{k-1}$ and covariance $P_{k-1}$, current observation $y_k$
\Ensure updated mean $m_k$ and covariance $P_k$
\State compute the predictive $\tilde{m}_k$ and covariance $\tilde{P}_k$ using \eqref{eq: predictive step}  \Comment{prediction step}
\State $t \gets 0$, $P_k^{(0)} \gets \tilde{P}_k$ \Comment{initialise optimiser at the prior}
\While{\texttt{not converged}}
  \State $t \gets t + 1$
  \State $P_k^{(t)} \gets \text{\texttt{L-BFGS}}\!\big(\phi_k(\,\cdot\,; P_k^{(t-1)})\big)$ \Comment{returns approx.\ solution to $\argmin_{P} \phi_k(P | P_k^{(t-1)})$}
  \State $m_k^{(t)}\gets m^\star(P_k^{(t)})$ \Comment{obtains the current best filter mean solution}
  \State Calculate \eqref{eq: pro linear1}, \eqref{eq: pro linear2} using $m_k^{(t)}$
\EndWhile
\State $P_k \gets P_k^{(t)}$ \Comment{updated covariance}
\State $m_k \gets m_k^{(t)}$ \Comment{updated mean}
\end{algorithmic}
\end{algorithm}

\paragraph{Guaranteed Improvement}

Because $\phi_k(\cdot ; P_k^{(t)})$ is a tight upper-bound on $\Phi_k$, with equality at $P=P_k^{(t)}$, each iteration is guaranteed to decrease the objective: 
$$
\Phi_k(P_k^{(t+1)}) \leq \phi_k(P_k^{(t+1)} ; P_k^{(t)}) \leq \phi_k(P_k^{(t)} ; P_k^{(t)}) = \Phi_k(P_k^{(t)}) 
$$
In practice we implement a sufficient number of iterations that $P_k^{(t)}$ no longer significantly changes when further iterations are performed.

\paragraph{Efficient Implementation via Cholesky Decomposition}

Direct optimisation over $P$ requires care since $P$ must remain a symmetric positive definite matrix.
Rather than implementing constrained optimisation over the manifold of such matrices, we instead parametrise $P$ using the lower-triangular Cholesky factor $L$ of the whitened matrix, i.e. $P = \tilde{P}_k^{1/2} L L^\top \tilde{P}_k^{1/2}$.
Our implementation applies \texttt{L-BFGS} to optimise with respect to $L$, whose domain is unconstrained.

\paragraph{Stopping Criterion}
For simplicity, we perform a fixed outer loop of length \texttt{n\_outer} for \mPrO. The stopping criterion for the inner \texttt{L-BFGS} loop is defined by 3 factors: (i) the hard ceiling of how many iterations to perform is determined by an \texttt{n\_inner} parameter; (ii) we threshold the norm of the gradient term $\lVert \nabla_P \phi(P,P')\rVert$ and upper bound this with \texttt{gtol}=$10^{-4}$; (iii) we check the ``stalling'' effect by checking if the relative improvement $\frac{\phi(\texttt{P}_{\texttt{curr}})-\phi(\texttt{P}_{\texttt{new}})}{\max\{1, \phi(\texttt{P}_{\texttt{curr}})\}} \leq \texttt{rtol}$, and stop if the above condition is achieved for \texttt{patience} number of iterations in a row. We set \texttt{rtol}$=10^{-7}$, \texttt{patience}$=3$. \texttt{n\_outer} and \texttt{n\_inner} can either be set manually or selected using Bayesian optimisation (Cf. \Cref{app: protocol}).

\section{Experimental Protocol}
\label{app: protocol}

\paragraph{Hardware Requirement}
Experiments were run locally on an Apple M2 Pro machine with 16 GB unified memory
using JAX/Metal. 

\paragraph{JAX implementation.}
Our experimental code is implemented in JAX. The code framework builds on the
third-party package \texttt{rebayes\_mini} (\url{https://github.com/gerdm/rebayes-mini/tree/main}), which provides a
lightweight implementation framework for recursive Bayesian filtering methods.
We use this package for shared filtering abstractions and baseline
implementations, and implement the method-specific updates, experimental
configurations, and evaluation scripts used in this paper on top of this
framework. 

\paragraph{Hyper-parameter selection} The hyper-parameters for different robust filters are chosen using the Bayesian Optimisation (BO) package \citep{nogueira2014}. 
We note that while \mPrO is hyper-parameter free, we can use BO to select the optimal number of iterative steps, similar to how the number of iterations in \mWangExtended and \mAgamenoniExtended can be tuned using BO. We argue that while these hyper-parameters are tunable, they are a consequence of the optimisation procedure and do not impact the final filtering distribution. For all methods, we perform hyper-parameter tuning by minimising mean-squared error using the first trial of the data-generating distribution. 

\subsection{Tracking an Object in 2D}
\label{app: tracking}

For each method, the positional \ac{MSE} is calculated as
$$
\mathsf{MSE} = \sum_{k=1}^T \sum_{i\in\{1, 2\}}(x_{k,i}- m_{k,i})^2 = \sum_{k=1}^T L_k^2,
$$
where $x_{k,i}$ is the $i$th coordinate of the true object location at time $k$, and $m_{k,i}$ is the filtering mean for the $i$th coordinate of the object location at time $k$.

\begin{figure}
    \centering
    \includegraphics[width=0.7\linewidth]{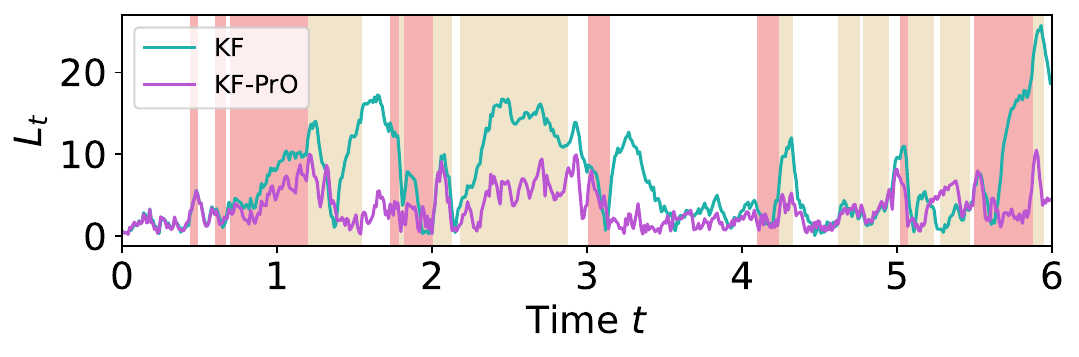}
    \caption{Example of the error $L_t = \lVert x_t - m_t \rVert$ for the location variable, as a function of time for \mkf and \mPrOKF in the \textbf{maneuver} type of misspecification. The background shading denotes the existence of systematic misspecification, with white background indicating no misspecification. We can see from this plot that \mPrOKF starts out similar to \mkf, but the propagation of overconfident predictions, and the consequent overconfidence priors for the next state leads to prolonged periods of time where \mkf contains much higher levels of error.}
    \label{fig:tracking maneuver}
\end{figure}

\paragraph{Generating a maneuvering object}
We consider a switching linear dynamical system, identical to Example 13.4.2 from \citep{pml2Book}, for tracking a maneuvering object.
The continuous state is
\[
x_t = (x_{1,t}, x_{2,t}, \dot{x}_{1,t}, \dot{x}_{2,t})^\top,
\]
where \((x_{1,t},x_{2,t})\) denotes the two-dimensional position and
\((\dot{x}_{1,t},\dot{x}_{2,t})\) denotes the corresponding velocity. Conditional on the discrete
mode \(m_t=k\), the state evolves according to
\[
p(x_t \mid x_{t-1}, m_t = k)
=
\mathcal{N}(x_t \mid F x_{t-1} + b_k, Q).
\]
The observations are linear-Gaussian,
\[
p(y_t \mid x_t)
=
\mathcal{N}(y_t \mid Hx_t, R),
\qquad
H = I,
\qquad
R = 10 I.
\]
The dynamics matrix is
\[
F =
\begin{pmatrix}
1 & 0 & \Delta & 0 \\
0 & 1      & 0 & \Delta \\
0 & 0      & 1 & 0 \\
0 & 0      & 0 & 1
\end{pmatrix},
\qquad
\Delta = 0.1,
\]
and the process noise covariance is
\[
Q = 0.1 I.
\]
The mode-dependent bias vectors are
\[
b_1 =
\begin{pmatrix}
0 \\ 0 \\ 0 \\ 0
\end{pmatrix},
\qquad
b_2 =
\begin{pmatrix}
-1.225 \\ -0.35 \\ 1.225 \\ 0.35
\end{pmatrix},
\qquad
b_3 =
\begin{pmatrix}
1.225 \\ 0.35 \\ -1.225 \\ -0.35
\end{pmatrix}.
\]
Thus the discrete mode controls the direction of the maneuver through the additive bias
\(b_k\). We use a persistent three-state Markov transition matrix with self-transition
probability \(0.99\),
\[
A =
\begin{pmatrix}
0.99  & 0.005 & 0.005 \\
0.005 & 0.99  & 0.005 \\
0.005 & 0.005 & 0.99
\end{pmatrix}.
\]
The comparison between \mkf and \mPrOKF is shown in \Cref{fig:tracking maneuver}, where we can see the propagation of overconfident predictions leads to prolonged periods of large errors obtained by \mkf, similar to the results shown in \Cref{fig:lorenz}.

\subsection{Lorenz96 Process}
\label{app: Lorenz}

\begin{figure}
    \centering
    \includegraphics[width=0.7\linewidth]{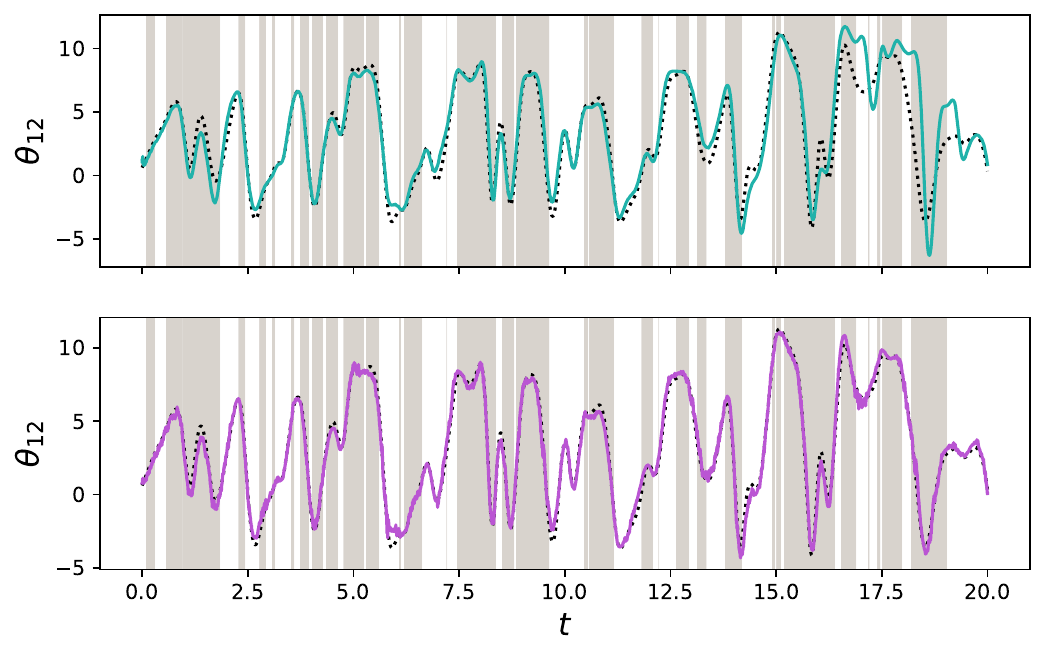}
    \caption{Tracing the mean of \mPrO and \mkfExtended against the ground truth state (black dotted). Even though \mkfExtended is able to capture the trajectory most of the time, it does not match \mPrO in accuracy, especially around the area $t=17.5$.}
    \label{fig:lorenz trace}
\end{figure}

The Lorenz96 process is a $d=60$-dimensional ordinary differential equation driven by some $\phi \in \mathbb{R}^d$. The canonical choice for $\phi$ is $8\cdot \mathbf{1}$, where $\mathbf{1}$ denotes a vector of all $1$s. For a fixed time interval $\Delta=0.002$, the data-generating process uses the following
\begin{align*}
    \frac{\theta_i^{t+\Delta} - \theta_i^t}{\Delta} = (\theta_{i+1}-\theta_{i-2})\theta_{i-1} - \theta_i + \phi_i. 
\end{align*}
Similar to the maneuver example, we set two discrete modes, with $\phi^{(1)} = 8\cdot \mathbf{1}$, $\phi^{(2)} \sim \mathcal{N}(8\cdot \mathbf{1}, \mathbf{I})$. The discrete model transition matrix is given by 
\begin{align*}
    A =
\begin{pmatrix}
0.995  & 0.005\\
0.005 & 0.995
\end{pmatrix}.
\end{align*}
The dynamics model is defined by the following 
\begin{align*}
    \theta_i^{t+\Delta} \sim \mathcal{N}\left(\theta_i^t + \Delta((\theta_{i+1}-\theta_{i-2})\theta_{i-1} - \theta_i + \phi^{(1)}_i), \Delta^2\right).
\end{align*}
and the state-space model is defined by $y^t\sim \mathcal{N}(\theta^t, 5\cdot \mathbf{I})$. The data-generating distribution shares the same measurement model Gaussian likelihood. We generate a total of $N=20,000$ steps. \Cref{fig:lorenz trace} illustrates the effect of a misspecified nonlinear state-space model on \mkfExtended. 

\end{document}